# Infrared spectra of both isomers of $CO_2$-CO in the $CO_2$ $\nu_3$ region

A.J. Barclay,[a] A.R.W. McKellar,[b] and  N. Moazzen-Ahmadi[a]

[a] *Department of Physics and Astronomy, University of Calgary, 2500 University Drive North West, Calgary, Alberta T2N 1N4, Canada*

[b] *National Research Council of Canada, Ottawa, Ontario K1A 0R6, Canada*

## Abstract

Extensive infrared spectra of the weakly-bound $CO_2$-CO dimer are observed in the carbon dioxide $\nu_3$ asymmetric stretch region ($\approx 2350$ cm$^{-1}$) using a tunable infrared OPO laser source to probe a pulsed slit jet supersonic expansion. Both C-bonded and O-bonded isomers are analyzed for the normal isotopologue as well as for $^{13}CO_2$-CO and $^{16}O^{13}C^{18}O$-CO, the latter being the first observation of an asymmetrically substituted form for which all values of $K_a$ are allowed. Combination bands involving the lowest in-plane intermolecular bending modes are also studied for both isomers. Weak bands near 2337 cm$^{-1}$ are assigned to $CO_2$ hot band transitions ($v_1$, $v_2^{l_2}$, $v_3$) = $(01^11) \leftarrow (01^10)$, yielding the splitting of the degenerate $CO_2$ $\nu_2$ bend into in-plane and out-of-plane components due to the presence of the CO. This splitting has rather different values for the C- and O- bonded isomers, 4.56 and 1.59 cm$^{-1}$, respectively, with the out-of-plane mode higher in energy than the in-plane for both cases.

## 1.  Introduction

The weakly-bound van der Waals dimer $CO_2$-CO was first observed by high-resolution microwave [1] and infrared [1-4] spectroscopy in the form of a planar T-shaped C-bonded structure ('isomer 1'), with $CO_2$ forming the 'top' of the T and CO forming the 'stem' and an intermolecular center of mass separation of about 3.9 Å. More recently, spectra of a second O-bonded isomer of $CO_2$-CO were observed [5]. This 'isomer 2' has a similar planar T-shaped structure, but with the CO flipped by 180º and a shorter intermolecular distance of about 3.6 Å. In addition to these gas-phase (supersonic jet) studies, infrared spectra of matrix-isolated $CO_2$-CO have been studied [6-8], but their relevance to the isolated dimer may be limited since interactions with the matrix itself are comparable in strength to the interaction between $CO_2$ and CO [8]. There have also been a number of theoretical studies of the $CO_2$ – CO interaction [9,10], including two very recent high level *ab initio* determinations of the four-dimensional potential surface [11,12].

Previously observed infrared spectra of $CO_2$-CO include those of isomer 1 in the $CO_2$ $\nu_3$ region ($CO_2$-CO and $CO_2$-$^{13}$CO) [1] and the CO stretch region ($CO_2$-CO and $C^{18}O_2$-CO) [2-34]. Isomer 2 was observed only in the CO stretch region ($CO_2$-CO and $C^{18}O_2$-CO) [4,5]. (Isotope numbers for C and O are 12 and 16 unless otherwise given.) In addition, combination bands involving low frequency intermolecular modes had been detected for both isomers in the CO stretch region [4,5]. In the present paper, we report extensive new data for $CO_2$-CO as observed in the $CO_2$ $\nu_3$ region. These include the first observations in this region of isomer 2 and of intermolecular combination bands, and the first observation of an asymmetrically substituted dimer ($^{16}O^{13}C^{18}O$-CO). Most importantly, we also detect spectra of both isomers corresponding to the weak $(\nu_1, \nu_2^{l_2}, \nu_3) = (01^11) \leftarrow (01^10)$ hot band of $CO_2$, and thus observe the splitting of the degenerate $CO_2$ $\nu_2$ bending mode into its in-plane and out-of-plane components due to the presence of the nearby CO molecule. This splitting, turns out to be rather different for isomers 1 and 2.

As mentioned, both $CO_2$-CO isomers have planar T-shaped equilibrium structures with $C_{2v}$ point group symmetry. The $a$-inertial axis coincides with the CO axis and is perpendicular to the $CO_2$ axis. Since rotation around this axis interchanges the carbon dioxide O-atoms, only levels with $K_a$ = even are allowed by nuclear spin statistics for ground vibrational state dimers containing $C^{16}O_2$ or $C^{18}O_2$. The $b$-axis is parallel to the $CO_2$ axis, so fundamental bands in the $CO_2$ $\nu_3$ (asymmetric stretch) region have $b$-type rotational selection rules. The $c$ axis is perpendicular to the plane of the dimer. Each isomer has four intermolecular vibrational modes: van der Waals stretch ($A_1$ symmetry), out-of-plane rock ($B_1$), and two in-plane bends ($B_2$). The in-plane bends can be described as a CO rock and $CO_2$ rock, or else as a geared and anti-geared bend.

## 2. Observed spectra

Spectra were recorded at the University of Calgary as described previously [13-15]. A pulsed supersonic slit jet expansion is probed using a rapid-scan optical parametric oscillator source. A typical expansion gas mixture contains about 0.08% carbon dioxide plus 0.3% carbon monoxide in helium carrier gas with a backing pressure of about 11 atmospheres. Wavenumber calibration is carried out by simultaneously recording signals from a fixed etalon and a reference gas cell containing room temperature $CO_2$. Spectral simulation and fitting are made using the PGOPHER software [16].

2.1.    Fundamental bands

Legon and Suckley [1] previously studied the fundamental band of $CO_2$-CO (and $CO_2$-$^{13}CO$) in the $CO_2$ $\nu_3$ region. Our current results for isomer 1 are similar, except that we observe the $K_a = 3 \leftarrow 4$ and $5 \leftarrow 4$ subbands in addition to the previously observed $1 \leftarrow 2$, $1 \leftarrow 0$, and $3 \leftarrow 2$ subbands. We also observe isomer 2, the O-bonded form, for the first time in this region. The spectrum is shown in Fig. 1, and results from its analysis are in Table 1. Note that the band origins of the two isomers are separated by only 0.230 cm$^{-1}$, so their spectra overlap almost completely. In contrast, the two isomer bands are separated by 7.950 cm$^{-1}$ in the CO stretch region [4,5], making it easier to detect and analyze the weaker isomer 2 spectrum. The $S$-reduced form of the asymmetric rotor Hamiltonian [16] is used here for consistency with Refs. 2-5. For isomer 1, the ground state parameters are fixed at previously determined [3,5] values which incorporate precise microwave data [1]. We assign 92 observed isomer 1 lines in terms of 117 transitions with $J$-values up to 12 (some lines are unresolved asymmetry doublets or other blends). There were 81 assigned transitions in Ref. 1. For the weaker isomer 2 fundamental, ground and excited state parameters are varied, and we assign 47 lines (59 transitions) belonging to the $K_a = 3 \leftarrow 4$, $1 \leftarrow 2$, $1 \leftarrow 0$, $3 \leftarrow 2$, and $5 \leftarrow 4$ subbands with $J$-values up to 9. The fundamental and combination (see below) bands were fitted simultaneously, with root mean square (rms) residuals of 0.00040 and 0.00036 cm$^{-1}$ for isomers 1 and 2, respectively.

We also study the $^{13}CO_2$-CO isotopologue, using isotopically enriched $^{13}CO_2$ in the expansion gas mix. The results are very similar to those of the normal species (see Fig. 2 and Table 2). For isomer 1, 77 lines (93 transitions) are observed in the $K_a = 3 \leftarrow 4$, $1 \leftarrow 2$, $1 \leftarrow 0$, and $3 \leftarrow 2$ subbands and fitted with an rms residual of 0.00079 cm$^{-1}$. For isomer 2, 38 lines (47 transitions) are observed in the same subbands and fitted with an rms residual of 0.00021 cm$^{-1}$.

For this isotopologue there are no previous data, and the ground state parameters are thus determined for the first time. They are very similar to those of the normal isotopologue, reflecting the fact that the $^{13}$C atom is located close to the dimer center of mass.

While using the $^{13}$CO$_2$ gas mix, a weak spectrum was detected in the 2266 cm$^{-1}$ region which we eventually realized must be the fundamental band of $^{16}$O$^{13}$C$^{18}$O-CO. For this species, there is no nuclear spin statistical effect, and all values of $K_a$ are possible. Thus the spectrum, shown in Fig. 3, looks different than any previously studied CO$_2$-CO spectrum. The fitted parameters are given in Table 3. For isomer 1, 79 lines (86 transitions) are assigned in the $K_a = 1 \leftarrow 2, 0 \leftarrow 1, 1 \leftarrow 0, 2 \leftarrow 1$, and $3 \leftarrow 2$ subbands and fitted with an rms residual of 0.00037 cm$^{-1}$. For isomer 2, 42 lines (47 transitions) are assigned in the same subbands and fitted with a residual of 0.00058 cm$^{-1}$. The natural abundance of $^{18}$O is 0.2%, and since there are two equivalent O atoms, the natural abundance of $^{16}$OC$^{18}$O is 0.4%. But the strength of the $^{16}$O$^{13}$C$^{18}$O-CO spectrum is spread among all $K_a$ values, not just even ones, so 0.2% remains its expected intensity ratio relative to $^{13}$CO$_2$-CO for the strongest lines in the band. We estimate that the strongest $^{16}$O$^{13}$C$^{18}$O-CO transitions (Fig. 3) actually had about 0.5% the intensity of the strongest $^{13}$CO$_2$-CO transitions (Fig. 2), though this is only approximate because laser power and jet expansion conditions may change between regions. Thus the observed strength of $^{16}$O$^{13}$C$^{18}$O-CO is roughly twice that which might be expected. A possible explanation for this strength may simply be that our enriched $^{13}$CO$_2$ gas sample is also somewhat enriched in $^{18}$O.

## 2.2.    Combination bands

In the CO stretch region, combination bands involving intermolecular modes have been studied previously for both isomers of CO$_2$-CO [4,5]. The two modes observed were the lower frequency in-plane bend ($B_2$ symmetry), with values of 24.34 (isomer 1) and 14.19 (isomer 2)

cm$^{-1}$, and the out-of-plane rock ($B_1$ symmetry) with values of about 43.96 (isomer 1) and 22.68 cm$^{-1}$ (isomer 2).

In the $CO_2$ $\nu_3$ region, we observe the in-plane bend for both isomers. This intermolecular mode and the $CO_2$ $\nu_3$ intramolecular mode both have $B_1$ symmetry, so their combination has $A_1$ symmetry and the combination bands originating in the ground state have $a$-type selection rules. Results for isomer 1 are illustrated in Fig. 4 for both $CO_2$-CO and $^{13}CO_2$-CO, with fitted parameters given in Tables 1 and 2. The resulting value for the in-plane bending mode frequency in the excited $CO_2$ $\nu_3$ state is 24.510 cm$^{-1}$ for $CO_2$-CO (and 24.501 cm$^{-1}$ for $^{13}CO_2$-CO), close to the value of 24.34 cm$^{-1}$ observed in the CO stretch region. The experimental value is also close to theoretical predictions of 24.44 and 24.45 cm$^{-1}$ (in the ground intramolecular state) from recent *ab initio* studies [11, 12]. (There are some inconsistencies in the intermolecular frequencies reported in Ref. 11; we use here values obtained from their Table 2). The very weak combination band of isomer 2 is not shown here, but its fitted parameters are included in Table 1 (this band was not observed for $^{13}CO_2$-CO). The resulting intermolecular mode frequency is 14.372 cm$^{-1}$. Again, we find that this value is quite close to that observed in the CO stretch region (14.174 cm$^{-1}$) [5] and to those from recent theoretical predictions [11, 12] (14.94 and 14.68 cm$^{-1}$).

Small perturbations are evident in the isomer 1 combination band spectra (Fig. 4) which clearly involve the upper state. For example, in the normal $CO_2$-CO spectrum, the $J_{KaKc} = 4_{04} \leftarrow 5_{05}$ and $4_{04} \leftarrow 3_{03}$ transitions both have a 'satellite' line which is 0.0106 cm$^{-1}$ higher and about 0.13 times the strength of the main line. Similarly, $6_{06} \leftarrow 7_{07}$ and $6_{06} \leftarrow 5_{05}$ show a satellite 0.0075 cm$^{-1}$ higher with about 0.33 times the strength. As well, the upper state levels $5_{24}$ and $7_{26}$ are split into two components separated by about 0.007 and 0.004 cm$^{-1}$, respectively. In the

$^{13}CO_2$-CO spectrum, perturbations only seem to affect upper state levels with $K_a = 2$ and $J < 5$. For example, the $3_{22}$ level is split into two components separated by 0.0136 cm$^{-1}$. Looking at the possibilities, it seems likely that the source of these perturbations is the excited intramolecular CO stretch mode of $CO_2$-CO combined with a highly excited ($\approx$230 cm$^{-1}$) intermolecular mode.

## 2.3. The $(01^11) \leftarrow (01^10)$ hot band

We observed a relatively weak but very interesting spectrum around 2337 cm$^{-1}$, shown in Fig. 5, whose explanation was not apparent at first. Eventually, we realized that it was due to $CO_2$-CO dimers with $CO_2$ undergoing the hot band transition $(v_1, v_2^{l2}, v_3) = (01^11) \leftarrow (01^10)$. One source of confusion here was that isomer 2 was stronger than isomer 1, opposite to all our other observed spectra. This point is discussed below in Sec. 3. The other source of confusion was that the spectrum is strongly influenced by Coriolis interactions affecting its upper and lower vibrational states. The degenerate $v_2$ bending mode of $CO_2$ splits into two components in the $CO_2$-CO dimer, and these in-plane and out-of-plane components are mixed by the Coriolis interaction. Observations of this effect in weakly-bound van der Waals complexes are surprisingly rare, probably because the bending fundamental of small linear molecules like $CO_2$ tend to lie in the more difficult far infrared region. But notably one such investigation was made by Ohshima et al. [17], who studied the $C_2H_2$-Ar dimer in the region of the $C_2H_2$ $v_5$ bending vibration ($\approx$730 cm$^{-1}$). We have now also observed the $(01^11) \leftarrow (01^10)$ hot band in $CO_2$-Ar [18] and $CO_2$-N$_2$ [19], which are simpler than $CO_2$-CO since there is only one isomer to deal with.

The in-plane (i-p) and out-of-plane (o-p) $CO_2$-CO vibrations corresponding to our lower state ($CO_2$ ($01^10$)) have $A_1$ and $B_1$ symmetries, respectively, for both isomers. In the upper state, $(01^11)$, we multiply by $B_2$ (for the $CO_2$ $v_3$ mode) and obtain $B_2$ and $A_2$ symmetries for the i-p and o-p modes, respectively. We expect $b$-type selection rules, just like the fundamental (Fig. 1). The i-p component transition is $B_2 \leftarrow A_1$ with $K_a$ = odd ← even subbands, and the o-p component is $A_2 \leftarrow B_1$ with $K_a$ = even ← odd subbands. We also expect strong Coriolis mixing between the i-p and o-p modes characterized by a matrix element

$$\langle \text{i-p}, J, k \mid H \mid \text{o-p}, J, k \pm 1 \rangle = \tfrac{1}{2}\ \xi_b \times [J(J + 1) - k(k \pm 1)]^{\frac{1}{2}},$$

where $k$ is signed $K_a$, and $\xi_b$ is the $b$-type Coriolis interaction parameter, related to the dimensionless Coriolis parameter zeta by the relation $\xi_b = 2B\zeta$, where $B$ is the $B$ rotational constant and $\zeta$ can take values between zero (no coupling) and unity (complete coupling).

The details of our analysis are given in Table 4, and illustrated by the simulated spectra in Fig. 5. The stronger isomer 2 spectrum is also the one most affected by Coriolis mixing. One result is that the prominent $Q$-branch of the i-p $K_a$ = 1 ← 0 subband ($\approx$2337.4 cm$^{-1}$) is compact and runs backwards, with higher $J$ transitions at lower wavenumbers. In contrast, the $Q$-branch of the o-p $K_a$ = 0 ← 1 subband ($\approx$2336.5 – 2336.8 cm$^{-1}$) runs in the normal direction and is quite spread out. It turns out that the i-p and o-p modes of isomer 2 are split by about 1.58 cm$^{-1}$, with i-p lower in energy, and that the Coriolis mixing is virtually complete ($\zeta \approx 1$). Once this is established, the spectrum can be very well fitted with rotational constants that are similar to those already known for isomer 2. We assigned 101 lines (109 transitions) in all subbands from $K_a$ = 2 ← 3, to 3 ← 2 (recall that odd ← even subbands are in-plane and even ← odd are out-of-plane). The rms residual was 0.00036 cm$^{-1}$. In addition to the expected transitions within the i-p or o-p stacks, we observed many 'forbidden' transitions between stacks which become allowed due to the Coriolis mixing. They follow selection rules $\Delta K_a$ = 0, $\pm 2$, $\Delta K_c$ = 0, $\pm 2$, and help to accurately determine the i-p to o-p splitting.

The isomer 1 spectrum is weaker, but more 'normal' in structure, as shown for example by the $Q$-branch of the i-p $K_a = 1 \leftarrow 0$ subband at 2337.2 – 2337.3 cm$^{-1}$. The o-p component was difficult to assign because of its weakness. This weakness and the 'normal' band structure are explained to the fact that the splitting between i-p and o-p components, about 4.6 cm$^{-1}$, is much larger here than for isomer 2. This means that there is less population in the o-p mode (hence the weakness) and less Coriolis interaction between modes (hence the normality). We assigned 77 lines (85 transitions) in all subbands from $K_a = 2 \leftarrow 3$ to $3 \leftarrow 2$, and the rms residual of the fit was 0.00034 cm$^{-1}$. But only 18 of these transitions were for the weak o-p mode, and hence it was necessary to fix the $B$ and $C$ parameters for this mode at ground state values. Alternatively, a similar fit could be obtained by varying $B$ and $C$ and fixing the Coriolis parameters, $\xi_b$, but we think the result given in Table 4 is more realistic. 'Forbidden' transitions between i-p and o-p levels were not observed for isomer 1 since the Coriolis mixing is less strong than for isomer 2, and as a result the splitting of the two modes is less precisely determined in this case.

The lower state of the hot band transitions corresponds to the first excited $CO_2$ bending state, $(v_1, v_2^{l2}, v_3) = (01^10)$. This is located at 667.380 cm$^{-1}$ for the free $CO_2$ molecule, but the present spectra do not tell us how much it is shifted in $CO_2$-CO. In Table 4, we call these unknown shifts X and Y for isomers 1 and 2, respectively. They are likely to be smaller than a few cm$^{-1}$, and direct observation of the $\nu_2$ fundamental in the 667 cm$^{-1}$ region will be required to determine them. However, the present spectra *do* tell us how much the $(01^11) \leftarrow (01^10)$ hot band origins of the dimer shift relative to free $CO_2$. These shifts are equal to +0.239 and +0.469 cm$^{-1}$ for the i-p modes of isomers 1 and 2 respectively. The o-p mode shifts are very similar, since the i-p to o-p splittings are similar in the upper and lower states, $(01^11)$ and $(01^10)$. These hot band shifts are similar to, but slightly larger than, those for the fundamental, which are +0.211 and

+0.441 cm$^{-1}$. Table 5 gives a summary of band origin shifts determined here; note that the current shifts in the $CO_2$ $\nu_3$ region are considerably smaller in magnitude than those in the CO stretch region [4,5].

### 3. Discussion and conclusions

The observed splitting of the $CO_2$ $\nu_2$ mode into i-p and o-p components is very different for the two $CO_2$-CO isomers: 4.56 cm$^{-1}$ for isomer 1 (C-bonded) and 1.59 cm$^{-1}$ for isomer 2 (O-bonded). This difference could perhaps be rationalized by noting that isomer 1 is more strongly bound than isomer 2 (calculated zero point binding energies are 287 and 222 cm$^{-1}$, respectively [12]). But this simple explanation is somewhat undermined by the fact that the $CO_2$ $\nu_3$ vibrational shift is actually larger for isomer 2 than for isomer 1 (Table 5). For comparison, in the case of $CO_2$-Ar the $\nu_2$ splitting is 0.877 cm$^{-1}$ [18], less than those of either $CO_2$-CO isomer. In the case of $CO_2$-N$_2$, the $\nu_2$ splitting is 2.307 cm$^{-1}$ [19], intermediate between the two $CO_2$-CO isomers, as might be expected.

The $\xi_b$ parameters (Table 4) give values around 1.00 to 1.02 for the dimensionless Coriolis parameter $\zeta$, similar to those determined for $CO_2$-Ar and $CO_2$-N$_2$. The maximum value of $\zeta$ should normally be unity within the harmonic approximation, but in the analogous case of $C_2H_2$-Ar, a $\zeta$-value of 1.52 was determined and this anomaly was ascribed to large amplitude intermolecular motions [17]. This difference between $CO_2$-CO (or $CO_2$-Ar) and $C_2H_2$-Ar is consistent with the fact that the latter is indeed very nonrigid, as shown by its very low energy intermolecular bending mode (6.2 cm$^{-1}$) [20] as compared to $CO_2$-CO (24.8 or 14.2 cm$^{-1}$).

$CO_2$-CO isomer 2 is about 65 cm$^{-1}$ less strongly bound than isomer 1, and was not detected in Refs. 1-3. Based on similar cases [14], we think that observation of isomer 2 requires using a supersonic expansion with helium carrier gas and a very dilute $CO/CO_2$ mixture. As the expansion begins, some dimers are 'trapped' in the higher energy state and subsequent collisions with helium are inefficient at promoting relaxation into the lower energy state. Thus the exact relative strength of the isomer spectra probably depends on details of the gas mixture and supersonic expansion conditions. (There could also be modest differences in infrared transition probabilities.) For the fundamental bands observed here (Fig. 1), isomer 1 is approximately 3.5 times stronger than isomer 2, and in our previous study in the CO stretch region, isomer 1 was about twice as strong.

However, as we have seen, isomer 2 is actually stronger than isomer 1 in the $(01^11) \leftarrow (01^10)$ spectrum of Fig. 5 (the exact ratio is a bit uncertain), even though the gas mixture and expansion conditions were similar to those of Fig. 1. How can we understand the prominence of isomer 2 in this hot band spectrum? We speculate that 'hot' dimers containing $CO_2$ in the $(01^10)$ state, formed early in the supersonic expansion and surviving until the end, are relatively more likely also to contain the 'hot' isomer 2. The reason could be that dimers which happen to undergo significant energy changing collisions which relax the $CO_2$ vibration from $(01^10)$ to $(000)$ are also more likely to be relaxed from isomer 2 to isomer 1.

In conclusion, spectra of both C- and O-bonded isomers of the $CO_2$-CO van der Waals complex have been studied in the $CO_2$ $\nu_3$ fundamental region ($\approx 2350$ cm$^{-1}$) using a tunable infrared OPO laser source to probe a pulsed slit jet supersonic expansion. Spectra were assigned for the normal isotopologue as well as for $^{13}C^{16}O_2$-CO and $^{16}O^{13}C^{18}O$-CO, the latter being the first observation of an asymmetrically substituted form for which all values of $K_a$ are allowed. In addition to the fundamental bands, combination bands involving the lowest in-plane intermolecular mode were also observed for isomer 1 (C-bonded) of $CO_2$-CO and $^{13}CO_2$-CO and for isomer 2 of $CO_2$-CO. A weak $CO_2$-CO spectrum near 2337 cm$^{-1}$ was assigned to the $CO_2$ hot band transition $(v_1, v_2^{l2}, v_3) = (01^11) \leftarrow (01^10)$. It analysis yielded the splitting of the degenerate $CO_2$ $v_2$ bending vibration into in-plane and out-of-plane modes due to the presence of the nearby CO molecule, which turns out to have values of 4.56 and 1.59 cm$^{-1}$ for isomers 1 and 2, respectively.

**Acknowledgements**

The financial support of the Natural Sciences and Engineering Research Council of Canada is gratefully acknowledged.

Table 1. Molecular parameters for $CO_2$ - CO (in $cm^{-1}$).[a]

| | Isomer 1 | | | Isomer 2 | | |
|---|---|---|---|---|---|---|
| | Ground state [3] | Excited state fundamental | Excited state $a$-type combination | Ground state [5] | Excited state fundamental | Excited state $a$-type combination |
| $\sigma_0$ | 0.0 | 2349.3539(1) | 2373.8637(1) | 0.0 [b] | 2349.5845(1) | 2363.9569(2) |
| $A$ | [0.3956764] | 0.392602(21) | 0.392334(52) | 0.397038(29) | 0.393928(33) | 0.356143(48) |
| $B$ | [0.06281439] | 0.0627934(28) | 0.0635355(97) | 0.074488(16) | 0.074430(14) | 0.076105(20) |
| $C$ | [0.05383985] | 0.0537620(19) | 0.0538731(81) | 0.062128(17) | 0.061992(14) | 0.062431(15) |
| $10^5\,D_K$ | [-1.167] | -0.921(87) | +1.89(31) | -2.42(14) | -2.42(14) [c] | |
| $10^5\,D_{JK}$ | [1.16239] | [1.16239] | [1.16239] | 2.838(68) | 2.838(68) [c] | |
| $10^7\,D_J$ | [2.2148] | [2.2148] | [2.2148] | 4.5(15) | 4.5(15) [c] | |
| $10^8\,d_1$ | [-4.280] | [-4.280] | [-4.280] | | | |
| $10^8\,d_2$ | [-2.512] | [-2.512] | [-2.512] | | | |

[a] Uncertainties (1σ) in parentheses are in units of the last quoted digit, and parameters in square brackets were fixed at the indicated values. Ground state parameters were fixed at previously determined values as shown.

[b] The energy of isomer 2 relative to isomer 1 is calculated to be about 65 $cm^{-1}$ [11,12].

[c] These excited state fundamental distortion parameters were constrained to equal those of the ground state.

Table 2. Molecular parameters for $^{13}CO_2$ - CO (in $cm^{-1}$).[a]

| | Isomer 1 | | | Isomer 2 | |
|---|---|---|---|---|---|
| | Ground state | Excited state fundamental | Excited state $a$-type combination | Ground state | Excited state fundamental |
| $\sigma_0$ | 0.0 | 2283.6863(1) | 2308.1876(2) | 0.0 | 2283.8997(3) |
| $A$ | 0.395634(30) | 0.391640(31) | 0.392031(45) | 0.397024(88) | 0.394025(84) |
| $B$ | 0.062297(10) | 0.0621619(80) | 0.063114(14) | 0.073806(46) | 0.073809(42) |
| $C$ | 0.053450(10) | 0.0533350(85) | 0.053438(16) | 0.061795(40) | 0.061413(45) |
| $10^5 D_K$ | [-1.167] | [-1.167] | [-1.167] | | |
| $10^5 D_{JK}$ | 1.088(85) | 0.886(68) | 1.21(13) | 1.27(31) | 1.83(33) |
| $10^7 D_J$ | 2.02(55) | 1.49(62) | 3.02(92) | 13.7(39) | -11.5(64) |
| $10^8 d_1$ | [-4.280] | [-4.280] | [-4.280] | | |
| $10^8 d_2$ | [-2.512] | [-2.512] | [-2.512] | | |

[a] Uncertainties (1$\sigma$) in parentheses are in units of the last quoted digit, and parameters in square brackets were fixed at the indicated values. The isomer 1 parameters $D_K$, $d_1$, and $d_2$ were fixed at their $^{12}CO_2$ – CO values (Table 1).

Table 3. Molecular parameters for $^{16}O^{13}C^{18}O$ - CO (in cm$^{-1}$).[a]

| | Isomer 1 | | Isomer 2 | |
|---|---|---|---|---|
| | Ground state | Excited state fundamental | Ground state | Excited state fundamental |
| $\sigma_0$ | 0.0 | 2266.1751(2) | 0.0 | 2266.3984(2) |
| $A$ | 0.37312(21) | 0.370321(82) | 0.37465(13) | 0.371852(65) |
| $B$ | 0.061271(24) | 0.061256(19) | 0.072760(26) | 0.072681(34) |
| $C$ | 0.052307(22) | 0.052217(21) | 0.060341(25) | 0.060129(29) |
| $10^5\,D_K$ | -3.8(44) | -1.95(89) | | |
| $10^5\,D_{JK}$ | 7.2(28) | 0.94(15) | 2.58(63) | 2.74(26) |
| $10^7\,D_J$ | 2.0(26) | 9.(22) | 9.5(27) | -3.4(42) |

[a] Uncertainties (1σ) in parentheses are in units of the last quoted digit.

Table 4. Molecular parameters for the $(01^11)$-$(01^10)$ hot band of $CO_2$ - CO (in cm$^{-1}$).[a]

| | $(01^10)$ i-p | $(01^10)$ o-p | $(01^11)$ i-p | $(01^11)$ o-p |
|---|---|---|---|---|
| **Isomer 1** | | | | |
| $\sigma_0$ | X | 4.557(40)+X | 2336.8720(1)+X | 2341.442(40)+X |
| $A$ | 0.396583(42) | 0.395515(32) | 0.393387(30) | 0.392398(68) |
| $B$ | 0.062855(32) | [0.06281][b] | 0.062689(40) | [0.06281][b] |
| $C$ | 0.053910(17) | [0.05384][b] | 0.0538513(65) | [0.05384][b] |
| $\xi_b$ | 0.12770(94) | | 0.12512(100) | |
| **Isomer 2** | | | | |
| $\sigma_0$ | Y | 1.5856(4)+Y | 2337.1023(2)+Y | 2338.6802(4)+Y |
| $A$ | 0.397497(48) | 0.402802(39) | 0.394458(44) | 0.393474(75) |
| $B$ | 0.074213(70) | 0.074553(45) | 0.073822(12) | 0.074864(31) |
| $C$ | 0.062074(34) | 0.062046(17) | 0.0620012(67) | 0.062095(16) |
| $10^5\,D_{JK}$ | 1.54(25) | | 2.13(24) | 4.78(44) |
| $\xi_b$ | 0.15044(19) | | 0.148846(60) | |

[a] Uncertainties (1σ) in parentheses are in units of the last quoted digit. Centrifugal distortion parameters were fixed to their ground state values from Table 1 except as shown. $\sigma_0$ is the term value (vibrational energy) of the state. X and Y are equal to the free $CO_2$ $\nu_2$ frequency (667.380 cm$^{-1}$) plus or minus unknown vibrational shifts which are unlikely to be more than a few cm$^{-1}$. i-p = in-plane; o-p = out-of-plane.

[b] $B$ and $C$ parameters for the weak isomer 1 o-p band were fixed at these ground state values.

Table 5. Vibrational shifts of $CO_2$-CO relative to the free $CO_2$ molecule (in $cm^{-1}$).

|  | Isomer 1 | Isomer 2 |
|---|---|---|
| $CO_2$-CO, CO stretch region [4,5] | +4.970 | -2.982 |
| $CO_2$-CO, $CO_2$ $\nu_3$ fundamental | +0.211 | +0.411 |
| $CO_2$-CO, $CO_2$ $\nu_3$ hot band [a] | +0.239 | +0.469 |
| $^{13}CO_2$-CO, $CO_2$ $\nu_3$ fundamental | +0.199 | +0.412 |
| $^{16}O^{13}C^{18}O$-CO, $CO_2$ $\nu_3$ fundamental | +0.204 | +0.427 |

[a] These are for the i-p modes; the o-p values are similar.

## Figure Captions

Fig. 1.   Observed and simulated spectra of $CO_2$-CO in the region of the $CO_2$ $\nu_3$ fundamental band. Simulated spectra are shown for the two isomers of $CO_2$-CO and for $CO_2$ dimer, with an effective rotational temperature of 2.2 K. The strongest lines belong to the $K_a = 1 \leftarrow 0$ $Q$-branch of isomer 1 at 2349.7 $cm^{-1}$. Here and in the other spectra shown, blank regions correspond to known lines of $CO_2$ monomer and $CO_2$-He dimer. Observed peaks at 2349.17 and 2349.47 $cm^{-1}$ are believed to be due to larger clusters.

Fig. 2.   Observed and simulated spectra of $^{13}CO_2$-CO in the region of the $CO_2$ $\nu_3$ fundamental band, recorded using an enriched $^{13}CO_2$ sample.

Fig. 3.   Observed and simulated spectra of $^{16}O^{13}C^{18}O$-CO in the region of the $CO_2$ $\nu_3$ fundamental band, recorded using an enriched $^{13}CO_2$ sample. The spectrum has a different appearance than those in Figs. 1 and 2 because all values of $K_a$ are allowed in the ground and excited states of this asymmetrically substituted isotopologue.

Fig. 4.   Observed and simulated spectra of isomer 1 of $CO_2$-CO and $^{13}CO_2$-CO in the region of the combination band involving the lowest in-plane intermolecular bending mode. Note their almost identical appearances. The central peak (2373.85 $cm^{-1}$) in the top panel is the $K_a = 2 \leftarrow 2$ $Q$-branch, and the small peak to its left (2373.80 $cm^{-1}$) is the $K_a = 4 \leftarrow 4$ $Q$-branch. The $^{13}CO_2$-CO $K_a = 2 \leftarrow 2$ $Q$-branch in the lower panel is obscured by a $CO_2$ monomer line.

Fig. 5   Observed and simulated spectra of $CO_2$-CO corresponding to the $(\nu_1, \nu_2^{l2}, \nu_3) = (01^11) \leftarrow (01^10)$ hot band of $CO_2$. These bands are strongly affected by Coriolis interaction between the in-plane and out-of-plane components of the $CO_2$ $\nu_2$ bend, especially for isomer 2. The out-of-plane component of isomer 1 is very weak because it lies 4.5 cm$^{-1}$ above the in-plane component and thus has little population.

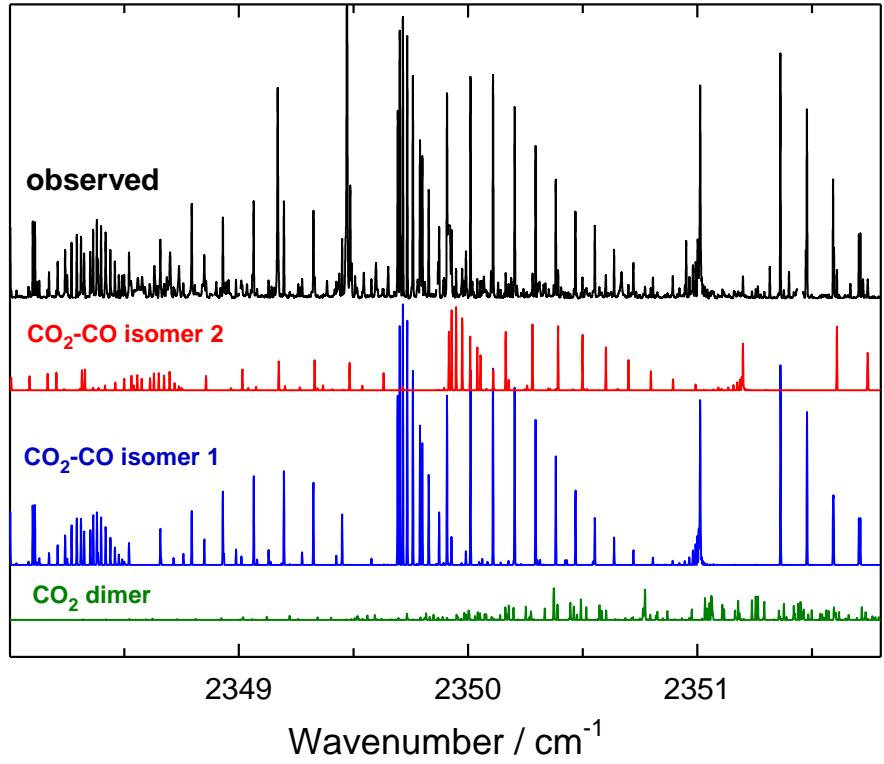

observed

CO₂-CO isomer 2

CO₂-CO isomer 1

CO₂ dimer

Wavenumber / cm⁻¹

Fig. 1

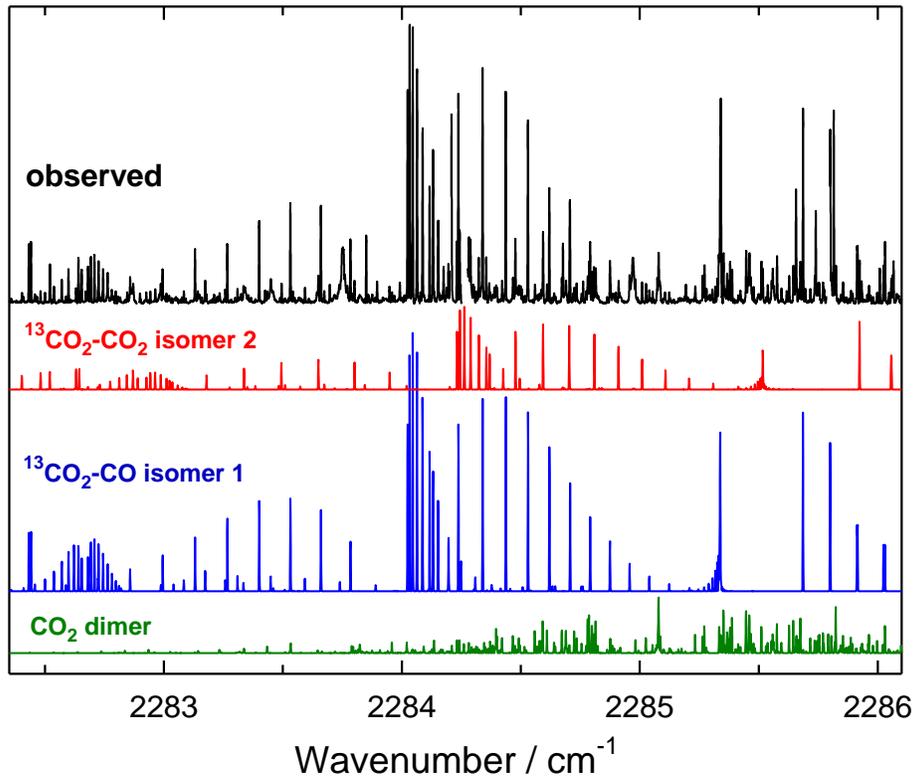

**observed**

**$^{13}CO_2$-$CO_2$ isomer 2**

**$^{13}CO_2$-$CO$ isomer 1**

**$CO_2$ dimer**

Wavenumber / cm$^{-1}$

Fig. 2

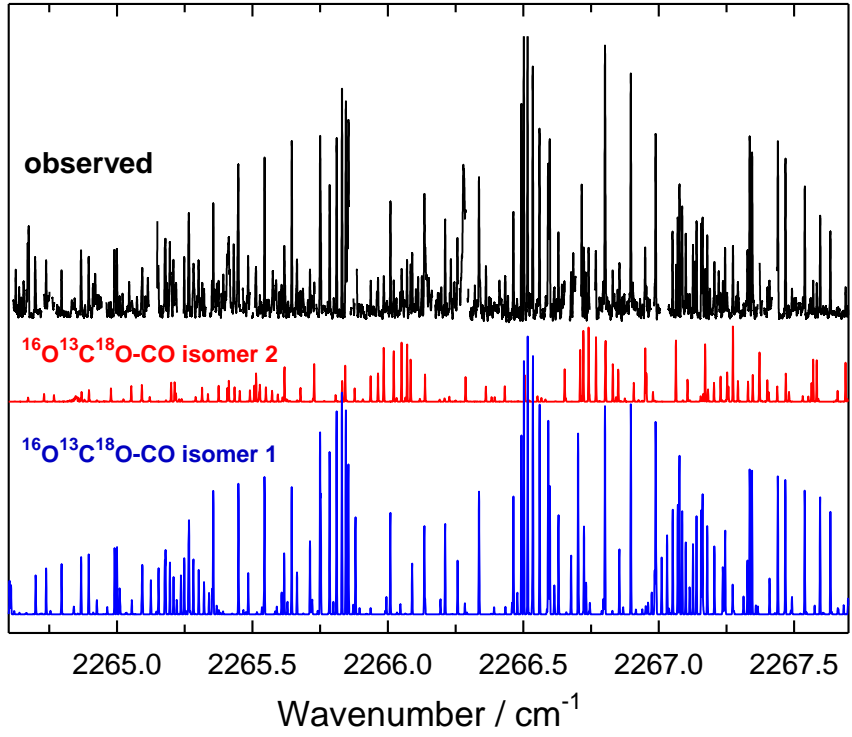

observed

$^{16}O^{13}C^{18}O$-CO isomer 2

$^{16}O^{13}C^{18}O$-CO isomer 1

Wavenumber / cm$^{-1}$

Fig. 3

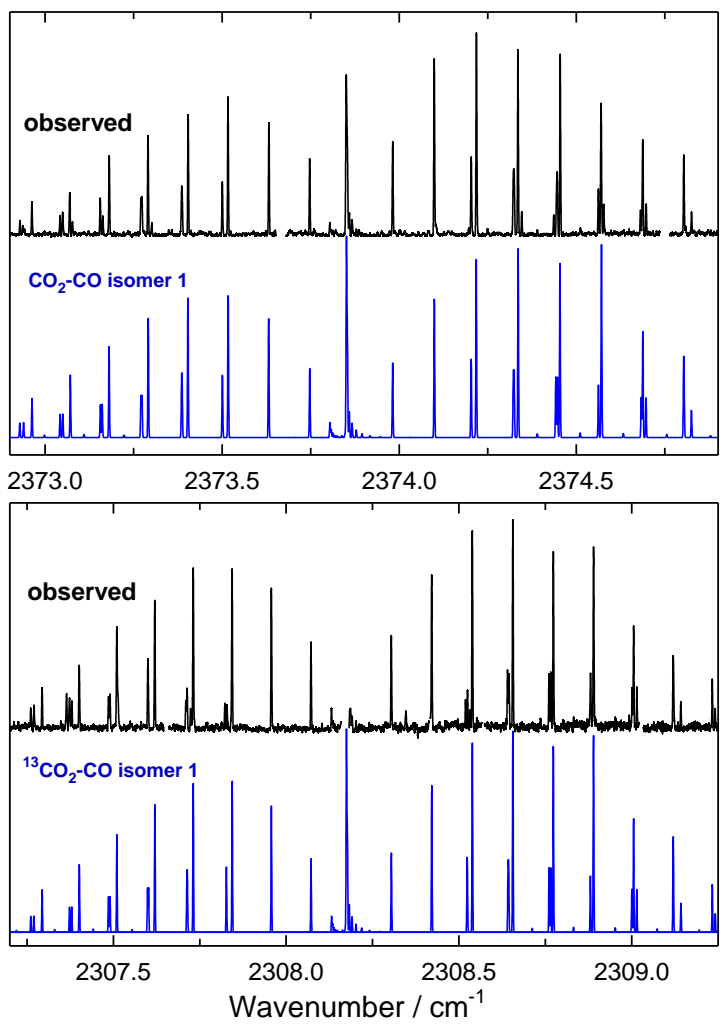

Fig. 4

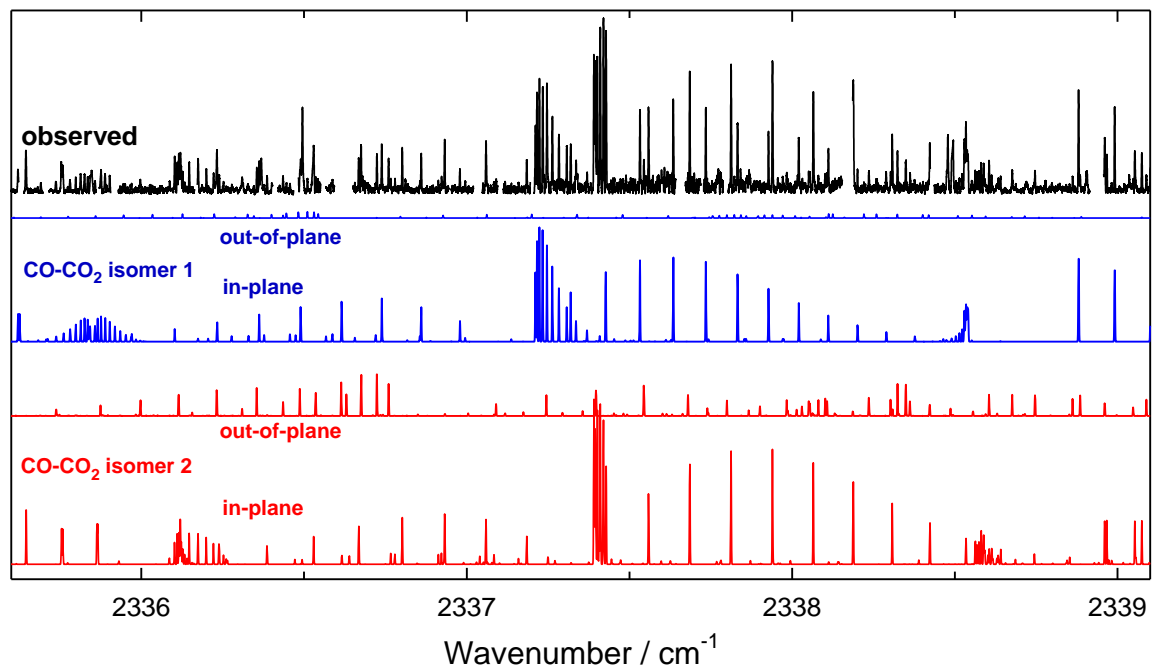

Fig. 5